\documentclass[12pt]{iopart}
\usepackage[utf8]{inputenc}
\usepackage[T1]{fontenc}
\usepackage{amsmath}
\usepackage{amsfonts}
\usepackage{amsthm}
\usepackage{amssymb}
\usepackage{mathtools}
\usepackage{bbm}
\usepackage{comment} 
\usepackage{marvosym}
\usepackage{graphicx}
\usepackage{xcolor}
\usepackage{appendix}
\usepackage{hyperref}

\newcommand{\ket}[1]{| #1 \rangle}

\newcommand{\kb}[2]{| #1 \rangle\hspace{-2pt}\langle #2 |}
\newcommand{\mean}[1]{\langle #1 \rangle}
\renewcommand{\tr}{\mathrm{Tr}}
\renewcommand{\Tr}{\mathrm{Tr}}

\newcommand{\s}{\{1,2,3\}}
\mathchardef\mhyphen="2D
\newcommand{\etac}{\eta_{\mathrm{crit}}}
\newcommand{\rhl}{{\rho_\lambda}}
\renewcommand{\i}{\mathrm{i}}

\newcommand{\1}{\mathbbm{1}}

\newcommand{\RR}{\mathbb{R}}
\newcommand{\nc}{\mathrm{\neg c}}
\newcommand{\mc}{\mathrm{c}}
\newcommand{\T}{\intercal}
\newcommand{\va}{(a_1,a_2,a_3)}
\newcommand{\vb}{(b_1,b_2,b_3)}

\begin{document}
\title{Entanglement Witnessing with Untrusted Detectors}
\author{Giuseppe Viola$^1$, Nikolai Miklin$^{1,2}$, Mariami Gachechiladze$^{3,4}$, Marcin Paw\l owski$^1$}
\address{$^1$ International Centre for Theory of Quantum Technologies (ICTQT), University of Gdansk, 80-309 Gda\'nsk, Poland}
\address{$^2$ Heinrich Heine University D\"usseldorf, Universit\"atsstra\ss e 1, 40225 D\"usseldorf, Germany}
\address{$^3$ Institute for Theoretical Physics, University of Cologne, Germany}
\address{$^4$ Department of Computer Science, Technical University of Darmstadt, Darmstadt, 64289 Germany}
\ead{giuseppe.viola@phdstud.ug.edu.pl}

\begin{abstract}
We consider the problem of entanglement detection in the presence of faulty, potentially malicious detectors. A common—and, as of yet, the only—approach to this problem is to perform a Bell test in order to identify nonlocality of the measured entangled state. However, there are two significant drawbacks in this approach: the requirement to exceed a critical, and often high, detection efficiency, and much lower noise tolerance. 
In this paper, we propose an alternative approach to this problem, which is resilient to the detection loophole and is based on the standard tool of entanglement witness. We discuss how the two main techniques to detection losses, namely the discard and assignment strategies, apply to entanglement witnessing. We demonstrate using the example of a two-qubit Bell state that the critical detection efficiency can be significantly reduced compared to the Bell test approach. 
\end{abstract}
\maketitle

\section{Introduction}
Entanglement of quantum states is one of the most cited and studied quantum phenomena~\cite{horodecki2009quantum}. This fact is largely explained by the simplicity of its formulation and a promising technological advancement that it offers, primary to the field of quantum cryptography~\cite{bennett1984quantum,gisin2002quantum}. At the same time, a variety of theoretical, experimental, and technological challenges are yet to be overcome before entanglement can find its real-world applications~\cite{liao2017satellite}. 

Among these challenges, perhaps the most crucial one is distribution of entangled particles among two or more laboratories that are far away from each other. It is rather clear that photons are the most natural, and essentially the only, candidates for such tasks. Entangled states of photons are easy to prepare (the first experiments date back to 1960s~\cite{kocher1967polarization,aspect1982experimental}), and, moreover, photons can be sent easily thorough optical fibers or simply the atmosphere~\cite{ursin2007entanglement}. However, there is a major downside in using photons as carriers of entanglement, namely a low efficiency of single-photon detectors~\cite{bedington2017progress}. 

The detection efficiency problem is most often discussed in the context of quantum cryptography or Bell experiments~\cite{bennett1984quantum,bell1964einstein}, where making the \emph{fair sampling assumption} can be unjustified. There, either an eavesdropper or a local hidden variable can gain control over the measurement devices, leading to insecure communication protocol or false nonlocality claims. However, it is often unjustified to assume the independence of detector's efficiency on the choice of measurement setting, which also affects simpler experiments such as entanglement detection.  

If the fair sampling is not assumed, non-detection events must be accounted for. Surprisingly, it is still possible to achieve unconditionally secure cryptography or a loophole-free Bell test demonstration for non-perfect detectors~\cite{pironio2009device,eberhard1993background}. However, such demonstrations are only possible for detection efficiencies above certain threshold values, which are often higher than the typical values of photodetectors used in today's experiments. 

Studying and lowering critical detection efficiency for quantum key distribution and Bell tests has been the subject of extensive research~\cite{brunner2007detection,vertesi2010closing,brown2021device,miklin2022exponentially,lukanowski2022upper}. At the same time, the problem of untrusted detectors in entanglement detection remains largely unexplored~\cite{skwara2007entanglement}. As of yet the only approach to this problem is to perform a Bell test, in which the detection loophole is closed~\cite{shalm2015strong,giustina2015significant}.
Apart from requiring high threshold detection efficiency, this approach can also be inapplicable to noisy entangled states that do not violate any of the known Bell inequalities. 

In this work, we take a different approach to the problem of entanglement detection with untrusted detectors. Instead of uplifting to a fully device-independent framework, we consider a scenario in which only the detection part of the measurement process, i.e., photon counting, is untrusted. The choice of basis, or more generally, the measurement setting, is assumed to be in a good control of the experimenter. This choice of assumptions is especially well-motivated for photonic experiments, where the part of the measurement device controlling measurement basis is much more studied than the detection part. Such scenarios in which only a part of the experimental apparatus is assumed to be characterized and trusted are often called \emph{semi-device-independent}~\cite{pawlowski2011semi}.   

For entanglement detection, we consider the common tool of entanglement witnessing~\cite{terhal2001family}. This method is universal, because it can be applied to any quantum state, and efficient, because it does not require performing the state tomography~\cite{horodecki2009quantum}. As the main technical contribution of this work, we discuss and analyze the discard and assignment strategies in entanglement witnessing with untrusted detectors.
We show that both of these approaches to dealing with imperfect detectors can be analyzed using semi-definite programming~\cite{vandenberghe1996semidefinite}. As an example, we apply our methods to two-qubit entanglement witnesses. Our results suggest that the critical detection efficiency of entanglement detection with untrusted detectors in the proposed semi-device-independent paradigm is significantly lower than that for Bell test experiments.

\section{Preliminaries}
We start by establishing notation and providing a few definitions required to present the results in the next sections. In this paper, we consider entanglement in bipartite systems, for which there is only one notion of separability and entanglement. Additionally, in our analysis, we consider qubit-qubit systems for the sake of simplicity. However, the results are directly generalizable to higher-dimensional systems, and we provide comments about this generalization whenever necessary. Finally, in this paper, we take all the observed efficiencies of all detectors to be the same and denote it by $\eta$. 

We take the standard definition of \emph{entanglement witness} as a linear Hermitian operator $W$, such that $\mean{W}_{\rho_s}\coloneqq \Tr[W\rho_s]\geq 0$ for all separable states $\rho_s$ and $\mean{W}_\rho<0$ for the entangled state $\rho$ under investigation.  In the presence of inefficient detectors, it also becomes important how the witness $W$ is measured in the experiment, the fact that was also pointed out in Ref.~\cite{skwara2007entanglement}. Typically, a witness $W$ is decomposed as a sum of tensor products of local observables.
In this work, we focus on two-qubit states for which the typical observables are the Pauli operators, i.e., we consider witnesses $W$ of the form
\begin{equation}\label{eq:w_decomp}
    W=w_{0,0}\1\otimes\1+\sum_{i=1}^3 w_{i,0}\sigma_i\otimes\1+\sum_{j=1}^3 w_{0,j}\1\otimes\sigma_j+\sum_{i,j=1}^3 w_{i,j}\sigma_i\otimes\sigma_j,  
\end{equation}
with $w_{i,j}\in \RR$ being coefficients of the decomposition, and $\{\sigma_1,\sigma_2,\sigma_3\} = \{\sigma_x,\sigma_y,\sigma_z\}$ being the set of Pauli operators. For qudit systems, a possible set of observables are the Heisenberg-Weyl operators~\cite{asadian2016heisenberg}. 

To evaluate the expectation value $\mean{W}_\rho$ in Eq.~\eqref{eq:w_decomp}, we need to estimate each of the terms $\mean{\sigma_i \otimes \sigma_j}_\rho$ for which $w_{i,j}\neq 0$ for all $i,j\in\{1,2,3\}$, as well as \emph{the marginal} terms $\mean{\sigma_i \otimes \1}_\rho$ for $i\in\s$ and $\mean{\1 \otimes \sigma_j}_\rho$ for $j\in\s$.  Let us denote the outcomes of the parties' measurements as ``$+$'' and ``$-$'', in which case the expectation values are calculated as
\begin{align}\label{eq:sigma_mean_two}
     \mean{\sigma_i\otimes\sigma_j}_\rho &= p(+,+|i,j)+p(-,-|i,j) -p(+,-|i,j)-p(-,+|i,j),
\end{align}
for $i,j\in\s$, and 
\begin{align}\label{eq:sigma_mean}
\begin{split}
    \mean{\1\otimes\sigma_j}_\rho &= p^B(+|j)-p^B(-|j),\\
    \mean{\sigma_i\otimes\1}_\rho &= p^A(+|i)-p^A(-|i),
\end{split}
\end{align}
otherwise. In the above, we used the superscript $^A$ and $^B$ to denote the marginal probabilities. All of the above probabilities can be estimated by the respective frequencies of detectors' clicks. The key problem that we investigate in this work is how the no-click events affect the expectation value of $W$ and, consequently, entanglement detection.

The literature that addresses the detection inefficiency problem, primarily in the context of the Bell test, describes two primary methods for handling no-click events~\cite{czechlewski2018influence}. In the first approach, referred to as the \emph{discard strategy}, one simply ignores all the events where at least one of the detectors did not click (for estimation of joint probabilities). Mathematically, it means that the joint probabilities as well as marginal probabilities in Eqs.~(\ref{eq:sigma_mean_two},\ref{eq:sigma_mean}) are replaced by the probabilities, conditioned on the click events:
\begin{align}\label{eq:discard_def}
p(+,+|i,j) &\mapsto p(+,+|i,j,\mc^A,\mc^B),
\end{align}
for all $i,j\in\s$ and similarly for other outcomes. In the above, $\mc^A$ and $\mc^B$ denote the \emph{events} of Alice's and Bob's detectors clicking. The marginal probabilities are mapped as
\begin{align}\label{eq:discard_def_marg}
    \begin{split}
        p^A(+|i) &\mapsto p^A(+|i,\mc^A),\\
        p^B(+|j) &\mapsto p^B(+|j,\mc^B),
    \end{split}
\end{align}
for all $i,j\in\s$.
The same mapping of probabilities is performed when one assumes the fair sampling. However, in case of the discard strategy (in Bell tests) one takes into account the effect of losses by increasing the local bound of a Bell inequality~\cite{branciard2011detection,czechlewski2018influence}. As we show in the next section, this also applies to entanglement witnessing: for imperfect detectors, the expectation value of a witness with respect to separable states can take negative values. However, there is still a range of values of $\eta$ for which entanglement detection is possible. 
 A similar observation has been made in Ref.~\cite{skwara2007entanglement}, which was the first to consider the problem of detection loophole in entanglement witnessing.
 
The second common approach of dealing with detection inefficiencies in Bell tests is the \emph{assignment strategy} method, sometimes also called binning. There, for every no-click event, one assigns one of the outcomes, in our case ``$+$'' or ``$-$'', either randomly or deterministically. Mathematically, it means that the evaluated probabilities are mapped as 
\begin{align}\label{eq:prob_trans_assign}
\begin{split}
p(+,+|i,j) &\mapsto \eta^2p(+,+|i,j,\mc^A,\mc^B)+\eta(1-\eta)p(+,+|i,j,\mc^A,\nc^B)\\
&+(1-\eta)\eta p(+,+|i,j,\nc^A,\mc^B)+(1-\eta)^2p(+,+|i,j,\nc^A,\nc^B),
\end{split}
\end{align}
for all $i,j\in\s$, and similarly for the other outcomes. In the above, $\nc^A$, $\nc^B$ denote the no-click events. The marginal probabilities are mapped as
\begin{align}\label{eq:prob_trans_assign_marg}
\begin{split}
p^A(+|i) &\mapsto \eta p^A(+|i,\mc^A)+(1-\eta)p^A(+|i,\nc^A),\\
p^B(+|j) &\mapsto \eta p^B(+|j,\mc^B)+(1-\eta)p^B(+|j,\nc^B),
\end{split}
\end{align}
for all $i,j\in\s$.

A particular assignment is determined by the form of the probabilities, conditioned on the events $\nc^A$ and $\nc^B$. For instance, if Alice chooses to always output ``$+$'' if her detector does not click, then we have that $p(+,+|i,j,\nc^A,\mc^B)=p^B(+|j,\mc^B)$ and  $p(-,+|i,j,\nc^A,\mc^B)=0$, etc. Since for an entanglement witness of the form in Eq.~\eqref{eq:w_decomp} we need to estimate the expectation values rather than probabilities, it is more convenient to define an assignment as follows 
\begin{align}\label{eq:assign_str_def}
\begin{split}
  a_i & \coloneqq p^A(+|i,\nc^A)-p^A(-|i,\nc^A),\\
  b_j & \coloneqq p^B(+|j,\nc^B)-p^B(-|j,\nc^B),
\end{split}
\end{align}
for $i,j\in\{1,2,3\}$. 

In Bell tests, the assignment strategy leads to lowering the maximally achievable quantum value of the Bell inequality, but does not increase the local bound~\cite{czechlewski2018influence}. As we show in the next section, this is not true in general for entanglement witnessing, and some local assignments can lead to negative values.

We demonstrate our findings on the example of pure entangled two-qubit states, for which we use a notation 
\begin{equation}\label{eq:partial_state}
    \ket{\Psi_\theta} \coloneqq \sin(\theta)\ket{0,0}+\cos(\theta)\ket{1,1}, 
\end{equation}
with $\theta\in(0,\frac{\pi}{4}]$. The corresponding witness for $\ket{\Psi_\theta}$ reads
\begin{equation}\label{eq:partial_w}
W_\theta \coloneqq \cos(\theta)^2\1\otimes\1-\kb{\Psi_\theta}{\Psi_\theta}.
\end{equation}
The non-zero coefficients of the decomposition of $W_\theta$ into Pauli observables are $w_{0,0} = \frac{1}{2}\cos(2\theta)+\frac{1}{4}$, $w_{0,3}=w_{3,0} = \frac{1}{4}\cos(2\theta)$, $w_{1,1}=-w_{2,2}=-\frac{1}{4}\sin(2\theta)$, and $w_{3,3} = -\frac{1}{4}$. We refer to $\ket{\Psi_{\frac{\pi}{4}}}$ and $W_{\frac{\pi}{4}}$ as the Bell state and the Bell witness, respectively.

\section{Results}

We start with a general formulation of the detection efficiency problem in entanglement witnessing. Similarly to the situation in Bell tests,  we cannot rule out the possibility that a \emph{hidden variable} $\lambda$ gains control over the detectors and correlates their efficiencies with the source of quantum states. Mathematically, it means that the observed joint probability distribution, e.g., for the outcome $+,+$, decomposes as
\begin{align}\label{eq:hid_var}
    \begin{split}
        p(+,+|i,j,\mc^A,\mc^B) & =  \sum_{\lambda\in\Lambda}p(+,+,\lambda|i,j,\mc^A,\mc^B) = \sum_{\lambda\in\Lambda}\frac{p(+,+,\lambda,\mc^A,\mc^B|i,j)}{p(\mc^A|i)p(\mc^B|j)}\\
        & = \frac{1}{\eta^2}\sum_{\lambda\in\Lambda}p(+,+,\mc^A,\mc^B|i,j,\lambda)p(\lambda) \\
        & = \frac{1}{\eta^2}\sum_{\lambda\in\Lambda}p(+,+|i,j,\mc^A,\mc^B,\lambda)p(\mc^A|i,\lambda)p(\mc^B|j,\lambda)p(\lambda),
    \end{split}
\end{align}
where we consider for simplicity the situation in which the detection events are uncorrelated and independent of the measurement choices with respect to the \emph{observed} probability distribution, i.e., $p(\mc^A,\mc^B|i,j) = p(\mc^A|i)p(\mc^B|j)=\eta^2$. The key observation here is that the response functions of click events, $p(\mc^A|i,\lambda)$ and $p(\mc^B|j,\lambda)$, may depend on the measurement settings $i$ and $j$. At the same time, since the entanglement source may also be controlled by the hidden variable $\lambda$, the states $\rho_\lambda$, with respect to which the outcome probabilities $p(+,+|i,j,\mc^A,\mc^B,\lambda)$ are calculated, may also vary with $\lambda$. In what follows, we show how to account for such situations in both, the discard and the assignment strategies.

\subsection{Discard strategy}

In this section, we discuss two main questions regarding the discard strategy: given an entanglement witness, what is the minimal value that it can take for separable states for a given detection efficiency $\eta$, and what is the critical detection efficiency $\etac$ for which no entanglement detection is possible? We show below how to formulate these questions as optimization problems, which can be cast as semidefinite programming problems (SDPs)~\cite{vandenberghe1996semidefinite}.

Looking at the expansion of the observed probabilities in Eq.~\eqref{eq:hid_var}, we can realize that the probabilities of click events, $p(\mc^A|i,\lambda)$ and $p(\mc^B|j,\lambda)$, without loss of generality can be taken to be deterministic, i.e., $0$ or $1$, by considering a sufficiently large set $\Lambda$. As a next step, we define sets $\Lambda^A_i$ and $\Lambda^B_j$ as 
\begin{align}
    \Lambda^A_i &= \left\{\lambda\in\Lambda \;\Big\vert\; p(\mc^A|i,\lambda) = 1 \right\},\quad \Lambda^B_j = \left\{\lambda\in\Lambda \;\Big\vert\; p(\mc^B|j,\lambda) = 1 \right\},
\end{align}
for $i,j\in\s$. We also consider unnormalized density operators $\rhl$, such that $\Tr[\rhl] = p(\lambda)$, and $\frac{\rhl}{p(\lambda)}$ is the state in which the particles are prepared for the value $\lambda$ of the hidden variable. This allows us to write the mapped expectation values $\mean{\sigma_i\otimes\sigma_j}$ simply as 
\begin{equation}
    \mean{\sigma_i\otimes\sigma_j} \mapsto \frac{1}{\eta^2}\sum_{\lambda\in\Lambda^A_i\cap\Lambda^B_j}\mean{\sigma_i\otimes\sigma_j}_{\rho_\lambda},
\end{equation}
and the marginal expectation values as 
\begin{align}
    \begin{split}
        \mean{\1\otimes\sigma_j} & \mapsto \frac{1}{\eta}\sum_{\lambda\in\Lambda^B_j}\mean{\1\otimes\sigma_j}_{\rho_\lambda},\\
        \mean{\sigma_i\otimes\1} & \mapsto \frac{1}{\eta}\sum_{\lambda\in\Lambda^A_i}\mean{\sigma_i\otimes\1}_{\rho_\lambda},
    \end{split}
\end{align}
for $i,j\in\s$. We can now formulate the problem of finding the minimal value of $W$ as the following SDP,
\begin{subequations}\label{eq:sdp_disc}
\begin{align}
    \min_{\rho_\lambda}\quad & w_{0,0}+\frac{1}{\eta}\sum_{i=1}^3w_{i,0}\sum_{\lambda\in\Lambda^A_i}\mean{\sigma_i\otimes\1}_{\rho_\lambda}+\frac{1}{\eta}\sum_{j=1}^3w_{0,j}\sum_{\lambda\in\Lambda^B_j}\mean{\1\otimes\sigma_j}_{\rho_\lambda}\nonumber \\
    & +\frac{1}{\eta^2}\sum_{i,j=1}^3w_{i,j}\sum_{\lambda\in\Lambda^A_i\cap\Lambda^B_j}\mean{\sigma_i\otimes\sigma_j}_{\rho_\lambda}, \nonumber \\
    \begin{split}\label{eq:sdp_disc_b}
    \text{s.t.}\quad & \sum_{\lambda\in\Lambda^A_i}\Tr[\rho_\lambda] = \sum_{\lambda\in\Lambda^B_j}\Tr[\rho_\lambda] = \eta, \; \sum_{\lambda\in\Lambda^A_i\cap \Lambda^B_j}\Tr[\rho_\lambda] = \eta^2,\;\;\forall i,j\in\s,
    \end{split} \\
    & \rho_\lambda \geq 0,\quad \rho_\lambda^{\T_A}\geq 0, \;\forall \lambda \in \Lambda, \label{eq:sdp_disc_c} \\
    & \rho_{\text{observed}} \geq 0, \quad  \sum_{\lambda\in\Lambda}\Tr[\rho_\lambda]=1. \label{eq:sdp_disc_d}
\end{align}
\end{subequations}
In the above SDP, we have introduced an operator $\rho_{\text{observed}}$, which corresponds to the physically observed state, and is defined as
\begin{align}\label{eq:observed_state}
        \rho_{\text{observed}} &\coloneqq  \frac{\1\otimes\1}{4}+\frac{1}{\eta}\sum_{i=1}^3\left(\sum_{\lambda\in\Lambda^A_i}\mean{\sigma_i\otimes\1}_{\rho_\lambda}\right)\frac{\sigma_i\otimes\1}{4}+\frac{1}{\eta}\sum_{j=1}^3\left(\sum_{\lambda\in\Lambda^B_j}\mean{\1\otimes\sigma_j}_{\rho_\lambda}\right)\frac{\1\otimes\sigma_j}{4}\nonumber \\
        & +\frac{1}{\eta^2}\sum_{i,j=1}^3\left(\sum_{\lambda\in\Lambda^A_i\cap\Lambda^B_j}\mean{\sigma_i\otimes\sigma_j}_{\rho_\lambda}\right)\frac{\sigma_i\otimes\sigma_j}{4}.
\end{align}
The condition $\rho_{\text{observed}}\geq 0$ in Eq.~\eqref{eq:sdp_disc_d} requests that the observed statistics corresponds to some physical state, which otherwise could lead to the parties realizing that the behavior of their detectors is malicious. Although the objective function of the SDP in Eq.~\eqref{eq:sdp_disc} could also be written as $\mean{W}_{\rho_{\text{observed}}}$, we found that it is more instructive to give a full expansion of the witness in Eq.~\eqref{eq:sdp_disc}. Alternatively to defining the state $\rho_{\text{observed}}$ in Eq.~\eqref{eq:observed_state}, one can request an existence of some density operator, such that the experimentally observed expectation values can be explained by this state. This can be relevant, e.g., in the situation when the decomposition of the witness in Eq.~\eqref{eq:w_decomp} features non-orthogonal observables.

The conditions in Eqs.~\eqref{eq:sdp_disc_b} ensure that the events of photons being detected by Alice's and Bob's devices appear to be uncorrelated and occur with the same probability $\eta$ for all the measurement settings. The conditions in Eq.~\eqref{eq:sdp_disc_b} ensure that the operators $\rho_\lambda$ are positive semidefinite and separable, due to the positive-partial-transpose criterion~\cite{horodecki2009quantum}. For the higher-dimensional case, the separability condition can be enforced by a hierarchy of SDP relaxations~\cite{doherty2004complete}.  

	\begin{figure}
	    \centering
	    \includegraphics[height=5.5cm]{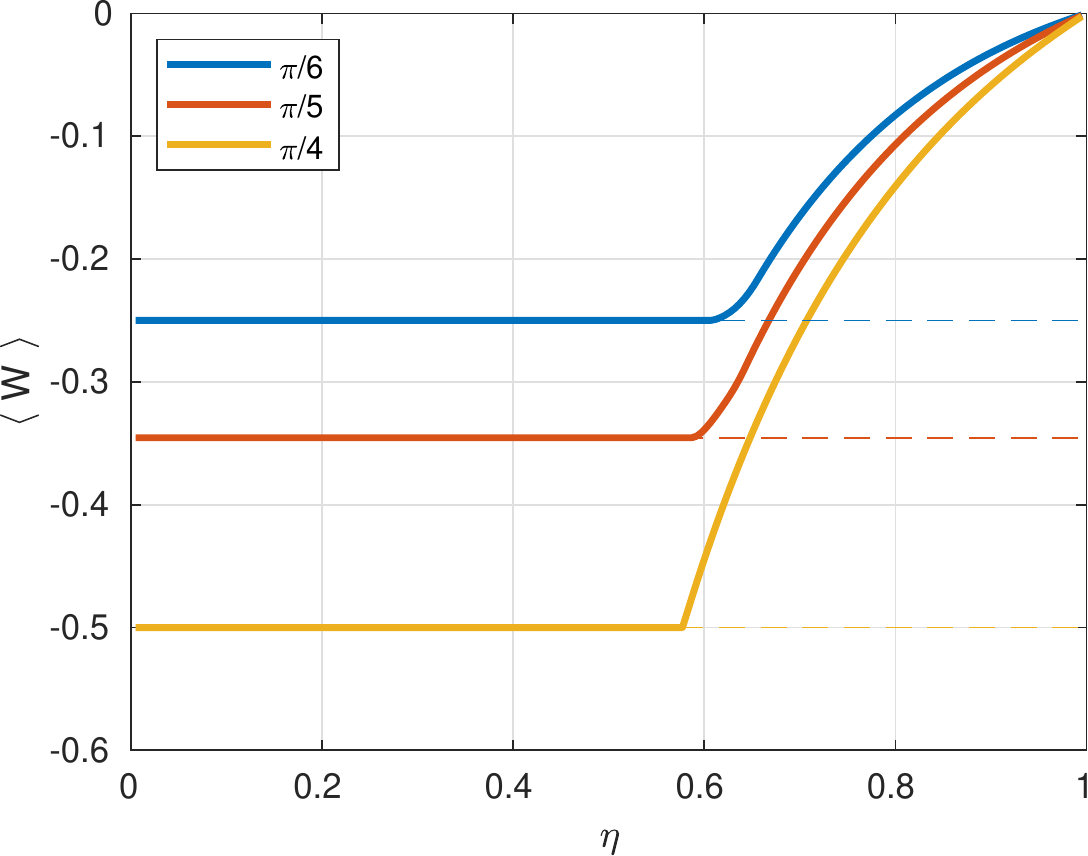}
	    \caption{Minimal values that the witness $\mean{W_\theta}$ can take in the discard strategy as a function of detection efficiency $\eta$.}
	    \label{fig:discard}
	\end{figure}

In Fig.~\ref{fig:discard} we demonstrate the solution of the SDP in Eq.~\eqref{eq:sdp_disc} for the witness $W_\theta$ in Eq.~\eqref{eq:partial_w} for $\theta\in\{\frac{\pi}{6},\frac{\pi}{5},\frac{\pi}{4}\}$. The values of $\eta$ for which $\mean{W_\theta}$ reaches its minimal value, shown by the dashed lines in Fig.~\ref{fig:discard}, is the critical detection efficiency. In~\ref{app:sec:discard}, we give an explicit solution for the Bell witness, and show that in this case the critical detection efficiency is $\frac{1}{\sqrt{3}}$. 
This value is significantly smaller than $0.83$, which is the critical detection efficiency of detecting the Bell state in Bell experiments~\cite{garg1987detector}.

\subsection{Assignment strategy}\label{sec:assign_main}
In this section, we discuss the assignment strategy to entanglement witnessing with untrusted detectors. As stated in the introduction, in Bell tests, the assignment strategy reduces the maximal possible violation of a Bell inequality while preserving the local hidden variable bound~\cite{branciard2011detection}. Additionally, there is no restriction on the particular choice of the assignment as long as it is performed locally by the parties. This, however, does not translate to the case of entanglement witnessing, as we show below. 

Let $\va$ and $\vb$, as defined by Eq.~\eqref{eq:assign_str_def}, be assignments chosen by the parties. Let us first assume that the behavior of the detectors is \emph{honest}, i.e., whenever the detectors click, the probabilities of outcomes, e.g., $p(+,+|i,j,\mc^A,\mc^B)$, correspond to a single state $\rho$ which is not in the control of the hidden variable. In that case, it is easy to observe that the transformation of probabilities due to the assignment strategy in Eq.~\eqref{eq:prob_trans_assign} can be equivalently captured by the following transformation of state $\rho$,
 \begin{align}\label{eq:trans_states}
 	&\rho \mapsto \eta^2\rho
 	+\eta(1-\eta)\rho^A \otimes \beta + (1-\eta)\eta  \alpha \otimes \rho^B +(1-\eta)^2 \alpha\otimes \beta,
 \end{align}
where $\rho^A$ and $\rho^B$ are the reduced states of Alice's and Bob's subsystems, and we have introduced the notation
\begin{equation}\label{eq:alpha_beta}
    \alpha \coloneqq \frac{\1}{2} +\frac{1}{2}\sum_{i=1}^3 a_i \sigma_i,\quad
    \beta  \coloneqq \frac{\1}{2} +\frac{1}{2}\sum_{i=1}^3 b_i \sigma_i.
\end{equation}

From the form of the state in Eq.~\eqref{eq:trans_states}, it is clear that as long as the initial state $\rho$ is separable and the operators $\alpha$ and $\beta$ in Eq.~\eqref{eq:alpha_beta} are positive semidefinite, the transformed state operator is also positive semidefinite and separable. Thus, we obtain a sufficient condition on the assignments for which no false detection of entanglement occurs:
\begin{equation}\label{eq:assign_cond}
    \sum_{i=1}^3 a_i^2\leq 1,\quad \sum_{i=1}^3 b_i^2\leq 1.
\end{equation}
Note that, a common deterministic assignment $p(+|i,\nc^A)=1$ for Bell tests, i.e., $a_i=1$ $\forall i\in \s$, does not satisfy the above constraint and can lead to a false detection of entanglement. In~\ref{app:sec:necess_assign}, we show that if the detection efficiencies of Alice's and Bob's detectors can be different, then there are values of them for which the above condition is also necessary. 

As an example of a good assignment, let us consider the Bell witness $W_{\frac{\pi}{4}}$ and the Bell state $\rho_{\frac{\pi}{4}} = \kb{\Psi_{\frac{\pi}{4}}}{\Psi_{\frac{\pi}{4}}}$. From the transformation in Eq.~\eqref{eq:trans_states}, we find that the expectation value of the witness equals to 
 \begin{equation}\label{eq:assign_witt_exp_bell}
 	\mean{W_{\frac{\pi}{4}}}_{\rho_{\frac{\pi}{4}}}=\frac{1}{2}\left(1-\eta^2-\eta-(1-\eta)^2 \Tr[\alpha^\T\beta]\right).
 \end{equation}   
It is clear, that a good assignment corresponds to taking  $\alpha$ and $\beta$ rank-$1$ and satisfying $\alpha^\T=\beta$. This can be achieved, e.g., by $\va = \vb=(1,0,0)$. One can also see from the above that entanglement of $\rho$ can be detected for $\eta>\frac{1}{2}$.

Now, we look at the general case of potentially malicious behavior of the detectors. This, in particular, means that the observed probabilities of outcomes, conditioned on click events, are given by Eq.~\eqref{eq:hid_var}. We start with the same argument that by considering a large enough set $\Lambda$, the probabilities of click events can be taken to be either $0$ or $1$. For the assignment strategy we would need to introduce an extra notation for subsets of $\Lambda$,
\begin{align}
    \overline{\Lambda}^A_i &= \left\{\lambda\in\Lambda \;\Big\vert\; p(\mc^A|i,\lambda) = 0 \right\},\quad \overline{\Lambda}^B_j = \left\{\lambda\in\Lambda \;\Big\vert\; p(\mc^B|j,\lambda) = 0 \right\},
\end{align}
which are just the complements of the sets $\Lambda^A_i$ and $\Lambda^B_j$. Using this notation, we can write the observed expectation values as
\begin{align}
\begin{split}
    \mean{\sigma_i\otimes\sigma_j} & \mapsto \sum_{\lambda\in\Lambda^A_i\cap\Lambda^B_j}\mean{\sigma_i\otimes\sigma_j}_{\rho_\lambda} + \sum_{\lambda\in\Lambda^A_i\cap\overline\Lambda^B_j}\mean{\sigma_i\otimes\1}_{\rho_\lambda}b_j\\
    & + \sum_{\lambda\in\overline\Lambda^A_i\cap\Lambda^B_j}\mean{\1\otimes\sigma_j}_{\rho_\lambda}a_i+ \sum_{\lambda\in\overline\Lambda^A_i\cap\overline\Lambda^B_j}\Tr[\rho_\lambda]a_ib_j,\\
\end{split}
\end{align}
for all pairs of $i,j\in\s$. 
The marginal expectation values are mapped as follows
\begin{align}
    \begin{split}
        \mean{\sigma_i\otimes\1} & \mapsto \sum_{\lambda\in\Lambda^A_i}\mean{\sigma_i\otimes\1}_{\rho_\lambda}+\sum_{\lambda\in\overline\Lambda^A_i}\Tr[\rho_\lambda]a_i,\\
        \mean{\1\otimes\sigma_j} & \mapsto \sum_{\lambda\in\Lambda^B_j}\mean{\1\otimes\sigma_j}_{\rho_\lambda}+\sum_{\lambda\in\overline\Lambda^B_j}\Tr[\rho_\lambda]b_j.
    \end{split}
\end{align}

We now formulate an SDP that determines the minimal value  of a witness for separable states given assignments $\va$ and $\vb$ and detection efficiency $\eta$. 
\begin{subequations}\label{eq:sdp_assign}
\begin{align}
    \min_{\rho_\lambda}\quad  & w_{0,0} + \sum_{i=1}^3w_{i,0}\left(\sum_{\lambda\in\Lambda^A_i}\mean{\sigma_i\otimes\1}_{\rho_\lambda}+\sum_{\lambda\in\overline\Lambda^A_i}\Tr[\rho_\lambda]a_i\right)\nonumber \\
    & \quad\;\;\, + \sum_{j=1}^3w_{0,j}\left(\sum_{\lambda\in\Lambda^B_j}\mean{\1\otimes\sigma_j}_{\rho_\lambda}+\sum_{\lambda\in\overline\Lambda^B_j}\Tr[\rho_\lambda]b_j\right)\nonumber\\
    & + \sum_{i,j=1}^3w_{i,j}\left(\sum_{\lambda\in\Lambda^A_i\cap\Lambda^B_j}\mean{\sigma_i\otimes\sigma_j}_{\rho_\lambda}+\sum_{\lambda\in\Lambda^A_i\cap\overline\Lambda^B_j}\mean{\sigma_i\otimes\1}_{\rho_\lambda}b_j\right.\nonumber\\
    & \left.\qquad\qquad\, +\sum_{\lambda\in\overline\Lambda^A_i\cap\Lambda^B_j}\mean{\1\otimes\sigma_j}_{\rho_\lambda}a_i+\sum_{\lambda\in\overline\Lambda^A_i\cap\overline\Lambda^B_j}\Tr[\rho_\lambda]a_ib_j,\right)\nonumber\\
    \begin{split}\label{eq:sdp_assign_b}
    \text{s.t.}\quad & \sum_{\lambda\in\Lambda^A_i}\Tr[\rho_\lambda] = \sum_{\lambda\in\Lambda^B_j}\Tr[\rho_\lambda] = \eta, \; \sum_{\lambda\in\Lambda^A_i\cap \Lambda^B_j}\Tr[\rho_\lambda] = \eta^2,\;\;\forall i,j\in\s,
    \end{split} \\
    & \rho_\lambda \geq 0,\quad \rho_\lambda^{\T_A}\geq 0, \;\forall \lambda \in \Lambda, \label{eq:sdp_assign_c} \\
    & \rho_{\text{observed}} \geq 0, \quad  \sum_{\lambda\in\Lambda}\Tr[\rho_\lambda]=1, \label{eq:sdp_assign_d}\\
    \begin{split}\label{eq:sdp_assign_e}
    & \sum_{\lambda\in\Lambda^A_i\cap\overline\Lambda^B_j}\mean{\sigma_i\otimes\1}_{\rho_\lambda} = (1-\eta)\sum_{\lambda\in\Lambda^A_i}\mean{\sigma_i\otimes\1}_{\rho_\lambda},\\
    &\sum_{\lambda\in\overline\Lambda^A_i\cap\Lambda^B_j}\mean{\1\otimes\sigma_j}_{\rho_\lambda} = (1-\eta)\sum_{\lambda\in\Lambda^B_j}\mean{\1\otimes\sigma_j}_{\rho_\lambda} \qquad \forall i,j,\in\s.
    \end{split}
\end{align}
\end{subequations}
where $\rho_{\text{observed}}$ is defined in Eq.~\eqref{eq:observed_state}. As in the case of the SDP for the discard strategy, the constraints in Eqs.\eqref{eq:sdp_assign_b} guarantee that the observed probabilities of click events are uncorrelated for Alice and Bob and are independent of their measurement settings. The conditions on $\rho_\lambda$ in Eq.~\eqref{eq:sdp_assign_c} and on $\rho_{\text{observed}}$ in Eq.~\eqref{eq:sdp_assign_d} are also motivated analogously to the discard strategy case.  Finally, the constraints in Eq.~\eqref{eq:sdp_assign_e} guarantee that the observed marginal state of Alice is independent of whether Bob detectors click or not, and analogously that Bob's marginal is independent of the behavior of Alice's detector.

	\begin{figure}
	    \centering
	    \includegraphics[height=5.5cm]{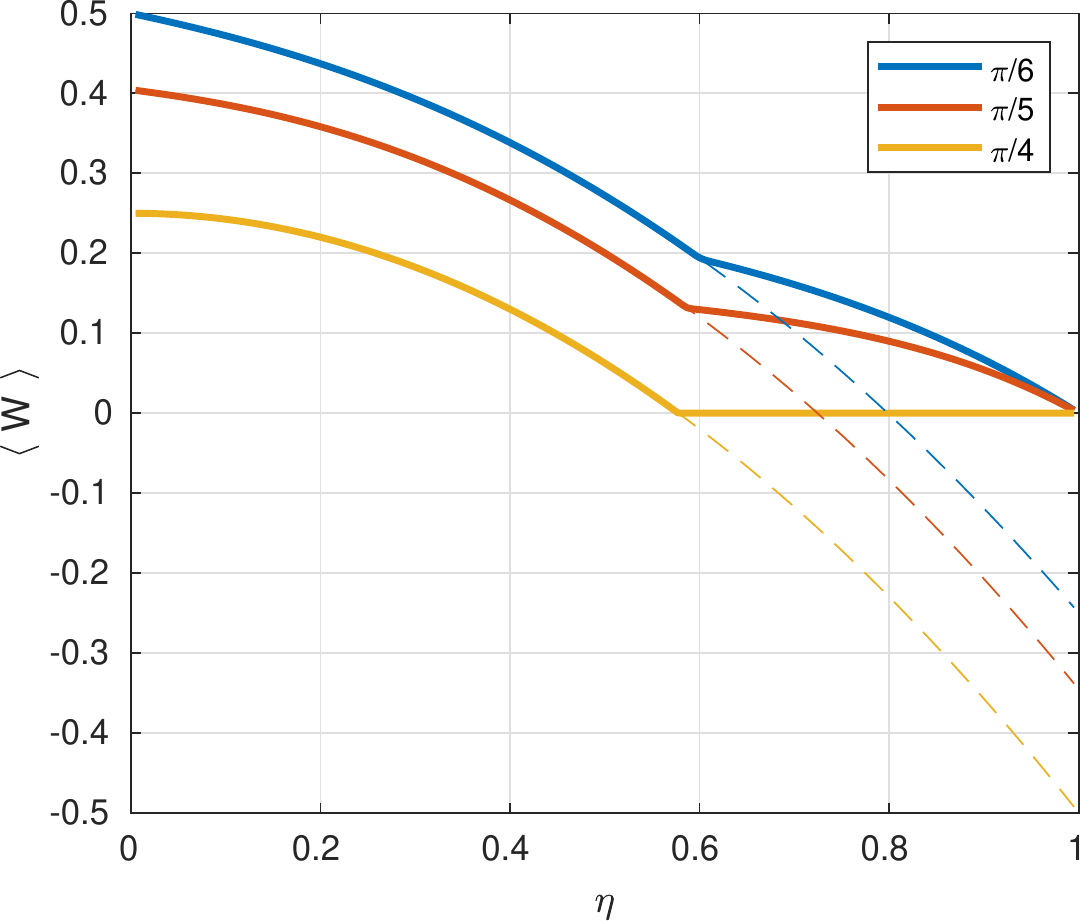} \quad \includegraphics[height=5.5cm]{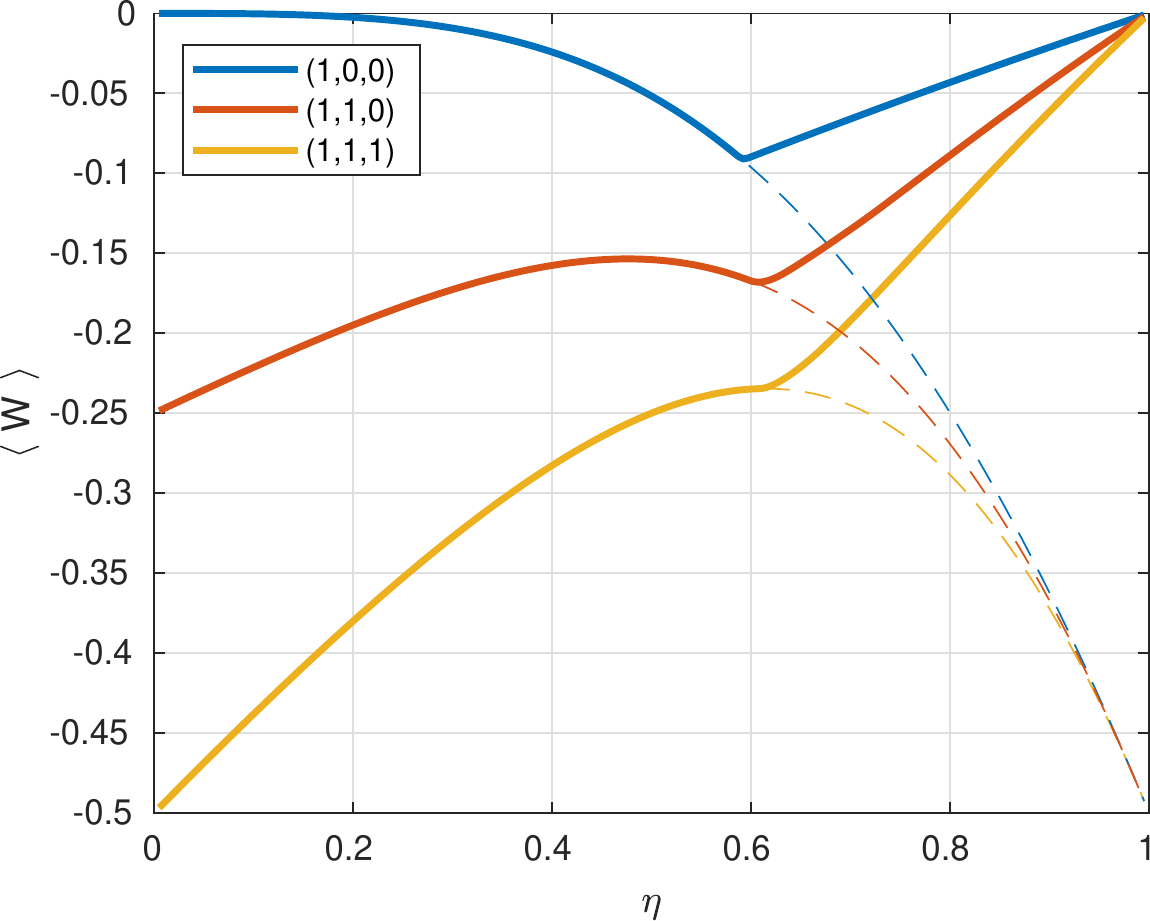}
	    \caption{Left: Minimal values of the witness $\mean{W_\theta}$ in the assignment strategy with $a_i=0$ and $b_j=0$, $\forall i,j,\in\s$ as a function of detection efficiency $\eta$ (solid lines) and the corresponding values of the witness for the entangled state $\ket{\Psi_\theta}$ (dashed lines). Right: Minimal values of the Bell witness for the assignments $\va \in \{(1,0,0),(1,1,0),(1,1,1)\}$ and $b_i = (-1)^{i+1}a_i$ $\forall i$ (solid lines), and the corresponding minimal values of the witness over entangled states (dashed lines).}
	    \label{fig:assignment}
	\end{figure}

In Fig.~\ref{fig:assignment} (left) we demonstrate the solution to the SDP in Eq.~\eqref{eq:sdp_assign} for the witness $W_\theta$ and the assignment $\va = \vb=(0,0,0)$. This particular assignment preserves the property $\mean{W}_{\rho_s}\geq 0$ for all separable states. In Fig.~\ref{fig:assignment} (left) by the dashed lines we depict values of the witness $W_\theta$ with respect to the corresponding state $\ket{\Psi_\theta}$. These values increase as $\eta$ decreases due to the assignment strategy. The points where the dashed lines intersect the solid lines are the critical detection efficiencies for the chosen assignment.  

In Fig.~\ref{fig:assignment} (right) we also compare different assignments for the Bell witness. One can see that the property of $\mean{W}_{\rho_s}\geq 0$ is not preserved for any of the selected assignments. At the same time, one can still detect entanglement if the expectation value with respect to an entangled state is lower than the calculated bound on $\mean{W_{\frac{\pi}{4}}}_{\rho_s}$ due to untrusted detectors, as we demonstrate in Fig.~\ref{fig:assignment}. Notably, the minimal expectation values that the witnesses can take with respect to entangled states (dashed lines) do not correspond to the Bell state.

The critical detection efficiency for the Bell witness in this calculation is again $\eta=\frac{1}{\sqrt{3}}$ as for the discard strategy. In~\ref{app:sec:assign} we give an explicit solution to the SDP in Eq.~\eqref{eq:sdp_assign} for the Bell witness. The value of $\mean{W_{\frac{\pi}{4}}}_{\rho_{\frac{\pi}{4}}}$ with respect to the Bell state can be calculated from Eq.~\eqref{eq:assign_witt_exp_bell} by taking $\alpha=\beta=\frac{\1}{2}$, and is equal to $\frac{1}{4}-\frac{3}{4}\eta^2$. 

\section{Conclusions and discussions}
In this paper, we discuss the problem of untrusted detectors in photonic experiments of entanglement detection. Even though the role of untrusted detectors is much more crucial for cryptographic applications and Bell tests, in this work we argue that malicious, or non-ideal, behavior of photodetectors can lead to false positive claims in simpler experiments of entanglement witnessing. We then analyze in detail the two main approaches to detection losses, namely the discard and the assignment strategies, and show that this analysis for a given entanglement witness can in both cases be cast as a semidefinite programming optimization problem. As an example, we analyze critical detection efficiencies for entanglement witnesses of pure two-qubit states. In particular, we show that the critical detection efficiency corresponding to the Bell state is $\frac{1}{\sqrt{3}}$ for the discard strategy. For the assignment strategy we could show that the same value of $\frac{1}{\sqrt{3}}$ can be attained, but the question whether this value can be reduced further by a suitable choice of an assignment is open.

On a more fundamental level, our work introduces a new type of semi-device-independent paradigm, the one in which only the detection part of measurement process is untrusted, while the measurement setting, e.g., the measurement basis, is assumed to be characterized. This is particularly relevant for the standard entanglement-based quantum key distribution (QKD), where the detectors are usually assumed to be fair. This assumption is difficult to justify because the most widely documented attacks against practical QKD are the detector blinding attacks, which boil down to gaining control of the detectors in the parties' devices. Additionally, dopant-level hardware Trojan attacks were demonstrated, which allow the manufacturer to place a malware in electronics, such as detector controllers, in a way impossible to detect by the user. We therefore believe that it is interesting to analyze security of QKD scheme in the introduced paradigm of untrusted detectors. As an additional motivation for this further work, one can notice that requirement on the detection efficiency reported in the current work is significantly lower than that in Bell test for the case of the Bell state.

\ack
We thank Dagmar Bru\ss{} for interesting discussions. 
This research was made possible by funding from QuantERA, an ERA-Net cofund in Quantum Technologies (www.quantera.eu) under project eDICT. 
We acknowledge the support by the Foundation for Polish Science (IRAP project, ICTQT, contract no.~MAB/2018/5, co-financed by EU within Smart Growth Operational Programme). 
This research was funded by the Deutsche Forschungsgemeinschaft (DFG, German Research Foundation), Project No.~441423094, and under Germany's Excellence Strategy – Cluster of Excellence Matter and Light for Quantum Computing (ML4Q) EXC 2004/1 – 390534769.

\begin{appendix}

\section{Discard strategy for the Bell witness}\label{app:sec:discard}
Here we provide an explicit solution to the SDP in Eq.~\eqref{eq:sdp_disc} for the Bell witness $W_{\frac{\pi}{4}}$. This solution, more precisely the minimal value that $\mean{W_{\frac{\pi}{4}}}$ can take for a given $\eta$, is shown in Fig.~\ref{fig:discard}. 

To describe the solution, i.e., to specify the operators $\rho_\lambda$ for each value of $\eta$, we need to introduce a few notations. Let $\Lambda$ be the set of all binary strings of length $6$. Each value of $\lambda$, specified by a string, specifies the probabilities of detectors' clicks with the first three bits corresponding to the measurement settings of Alice, and the second three bits corresponding to the measurement settings of Bob. For example, for $\lambda=(1, 0, 0, 0, 1, 0)$, we have that $p(\mc^A|1,\lambda)=1$ and $p(\mc^B|2,\lambda)=1$, while the other probabilities are zero. Clearly, this choice of the alphabet of $\lambda$ is sufficient for the problem in case of three measurement settings per party.  

Let us define the following states
\begin{align}\begin{split}
    \rho_{x,x} & \coloneqq \frac{1}{2}\left(\kb{+,+}{+,+}+\kb{-,-}{-,-}\right),\\
    \rho_{y,y} & \coloneqq \frac{1}{2}\left(\kb{+\i,-\i}{+\i,-\i}+\kb{-\i,+\i}{-\i,+\i}\right),\\
    \rho_{z,z} & \coloneqq \frac{1}{2}\left(\kb{0,0}{0,0}+\kb{1,1}{1,1}\right),
    \end{split}
\end{align}
which are separable states with the property that $\mean{\sigma_1\otimes\sigma_1}_{\rho_{x,x}} = \mean{\sigma_3\otimes\sigma_3}_{\rho_{z,z}} = 1$, and $\mean{\sigma_1\otimes\sigma_1}_{\rho_{y,y}} = -1$. Here, we used the notation $\ket{+\i}$ and $\ket{-\i}$ to denote the $+1$ and $-1$ eigenstates of $\sigma_2$. In terms of these states, the operators $\rho_\lambda$ in our solution are specified to be the following
    \begin{align}\label{app:eq:solution_rhos}
     \rho_{(0,0,0,0,0,0)} & = p_0\frac{\1\otimes\1}{4},\; \rho_{(1,1,1,1,1,1)} = p_1\frac{1}{3}(\rho_{x,x}+\rho_{y,y}+\rho_{z,z}),\\
        \rho_{(1,0,0,1,0,0)} & = \rho_{(1,0,1,1,1,0)} = \rho_{(1,1,0,1,0,1)} = p_2\rho_{x,x},\; \rho_{(1,0,0,1,1,1)} = \rho_{(1,1,1,1,0,0)} = p_3\rho_{x,x},\nonumber\\
             \rho_{(0,1,0,0,1,0)} & = \rho_{(0,1,1,1,1,0)} = \rho_{(1,1,0,0,1,1)} = p_2\rho_{y,y},\; \rho_{(0,1,0,1,1,1)} = \rho_{(1,1,1,0,1,0)} = p_3\rho_{y,y},\nonumber\\
            \rho_{(0,0,1,0,0,1)} & = \rho_{(0,1,1,1,0,1)} =\rho_{(1,0,1,0,1,1)} = p_2\rho_{z,z},\; \rho_{(0,0,1,1,1,1)} = \rho_{(1,1,1,0,0,1)} = p_3\rho_{z,z},\nonumber\\
            \rho_{(0,0,0,0,0,1)} & = \rho_{(0,0,0,0,1,0)} = \rho_{(0,0,0,1,0,0)} = \rho_{(0,0,1,0,0,0)} = \rho_{(0,1,0,0,0,0)} = \rho_{(1,0,0,0,0,0)} = p_4\frac{\1\otimes\1}{4},\nonumber
    \end{align}
    where the parameters $p_i\in \RR$, $i\in\{0,1,2,3,4\}$ will be specified later. 
 The rest of the operators $\rho_\lambda$ are taken to be zero-trace.  

The objective function of the SDP in Eq.~\eqref{eq:sdp_disc} can be calculated to be
\begin{align}
    \begin{split}\label{app:eq:bell_witt_disc}
        & \frac{1}{4}-\frac{1}{4\eta^2}\left(\sum_{\lambda\in\Lambda^A_1\cap\Lambda^B_1}\mean{\sigma_1\otimes\sigma_1}_{\rho_\lambda}-\sum_{\lambda\in\Lambda^A_2\cap\Lambda^B_2}\mean{\sigma_2\otimes\sigma_2}_{\rho_\lambda}+\sum_{\lambda\in\Lambda^A_3\cap\Lambda^B_3}\mean{\sigma_3\otimes\sigma_3}_{\rho_\lambda}\right)\\
        & = \frac{1}{4}-\frac{1}{4\eta^2}\left(p_1+9p_2+6p_3\right),
    \end{split}
\end{align}
while the constraints in Eq.~\eqref{eq:sdp_disc} correspond to 
\begin{align}\label{app:eq:constraints_ps}
    \begin{split}
        & p_1+5p_2+4p_3+p_4 = \eta,\\
        & p_1+3p_2+2p_3 = \eta^2,\\
        & p_0+p_1+9p_2+6p_3+6p_4 = 1.
    \end{split}
\end{align}
The observed state in Eq.~\eqref{eq:observed_state} can be easily found to be
\begin{align}\begin{split}
    \rho_{\text{observed}} = & \frac{\1\otimes\1}{4}+\frac{1}{4\eta^2}\left(\frac{p_1}{3}+3p_2+2p_3\right)(\sigma_1\otimes\sigma_1-\sigma_2\otimes\sigma_2+\sigma_3\otimes\sigma_3) \\
    = & \frac{\1\otimes\1}{4}\left(1-\frac{\frac{p_1}{3}+3p_2+2p_3}{\eta^2}\right)+\frac{1}{\eta^2}\left(\frac{p_1}{3}+3p_2+2p_3\right)\kb{\Psi_{\frac{\pi}{4}}}{\Psi_{\frac{\pi}{4}}}.
\end{split}\end{align}
From the above expression, it is clear that as long as $0\leq\frac{p_1}{3}+3p_2+2p_3\leq \eta^2$, the above density operator is positive semidefinite, and since this constraint is implied by Eqs.~\eqref{app:eq:constraints_ps}, these are the only constraints associated with the SDP.

Now, we can specify our solution to the original optimization problem in terms of the coefficients $\{p_i\}_{i=0}^4$.
For the case of $\eta>\frac{1}{\sqrt{3}}$, the minimal value which can be attained is $\frac{1}{4}-\frac{1}{4\eta^2}$, which is also clear from the expression in Eq.~\eqref{app:eq:bell_witt_disc}. This is achieved for the following values of the parameters,
\begin{equation}\label{app:eq:solution_ps_1}
    p_0=p_4=0,\; p_1 = \frac{3\eta^2-1}{2},\; p_2 = \frac{(1-\eta)^2}{2},\; p_3 = \frac{(1-\eta)(2\eta-1)}{2}.
\end{equation}
For the case of $\eta\leq \frac{1}{\sqrt{3}}$, the minimal value $-\frac{1}{2}$ can be reached. The corresponding values of the parameters are $p_1=0$ and
\begin{align}\label{app:eq:solution_ps_2}
    \begin{split}
        p_0 = (1-3\eta)^2,\; p_2=0,\; p_3=\frac{\eta^2}{2},\; p_4 = \eta-2\eta^2,\quad &\text{for}\; \eta\leq\frac{1}{3},\\
        p_0 = 0,\; p_2 = \frac{(1-3\eta)^2}{6},\; p_3 = \frac{6\eta-1-7\eta^2}{4},\; p_4 = \frac{1-3\eta^2}{6},\quad &\text{for}\; \eta\in\left(\frac{1}{3},\frac{1}{\sqrt{3}}\right].
    \end{split}
\end{align}

\section{Necessity of the condition in Eq.~\eqref{eq:assign_cond} for the assignment strategy in case of different detectors' efficiencies}\label{app:sec:necess_assign}
Here, we show that if the detection efficiencies can be different for Alice and Bob, then the condition in Eq.~\eqref{eq:assign_cond} is also necessary. More precisely, there are values of the detection efficiencies $\eta^A$ and $\eta^B$ for which this condition is necessary.

First, consider the case of $\eta^A=0$ and $\eta^B=1$. For any separable state ${\rho}_s$ and a witness $W$, the following must hold true, $\tr(W \alpha \otimes \rho^B_s)\geq 0$, where $\rho^B_s = \tr_A[\rho_s]$. If we take the Bell state witness $W_{\frac{\pi}{4}}$, the condition further simplifies to $\tr[\alpha^\T \rho_s^B]\leq 1$ for all states $\rho^B_s$. Inserting in the latter a particular state of the form, 
\begin{equation}
    \rho^B_s=\frac{\1}{2}+\frac{1}{2}\sum_{i=1}^3 \frac{a_i}{\sqrt{\sum_{j=1}^3a_j^2}}\sigma_i,
\end{equation}
directly results into the condition $\sum_{i=1}^3 a_i^2\leq 1$. The same way, we can prove the necessity of the constraint on $b_i$'s for the situation when $\eta^A=1$ and $\eta^B=0$.

\section{Assignment strategy for the Bell witness}\label{app:sec:assign}
The solution to the SDP in Eq.~\eqref{eq:sdp_assign} for the assignment strategy can be taken in the same form as in the case of the discard strategy, as described in~\ref{app:sec:discard}. In particular, one can take the states $\rho_\lambda$ as in Eq.~\eqref{app:eq:solution_rhos}, which leads to the constraints of the SDP to take the form in Eq.~\eqref{app:eq:constraints_ps}. The particular solution in terms of the parameters $p_0,p_1,p_2,p_3$ and $p_4$ is also the same as in Eq.~\eqref{app:eq:solution_ps_1} for $\eta>\frac{1}{\sqrt{3}}$ and as in Eq.~\eqref{app:eq:solution_ps_2} for $\eta\leq\frac{1}{\sqrt{3}}$. The only difference with the case of the discard strategy is the optimal value of the objective function, which is equal to $0$ for $\eta>\frac{1}{\sqrt{3}}$ and $\frac{1}{4}-\frac{3}{4}\eta^2$ for $\eta\leq\frac{1}{\sqrt{3}}$.

\end{appendix}
\clearpage

\section*{References}
\bibliographystyle{unsrt}
\bibliography{bibliography}

\end{document}